\documentclass[nofootinbib,showpacs,preprintnumbers,
amsmath,amssymb,superscriptaddress]{article}
\usepackage{a4wide}
\usepackage{epsfig}

\begin{document}

\title{FORWARD-BACKWARD CHARGE ASYMMETRY AT VERY HIGH ENERGIES}

\author{
  B.I.~Ermolaev $^{a\, b}$,    
  M.~Greco $^c$, 
  S.M.~Oliveira $^a$ and  
  S.I.~Troyan $^d$\\
  { \em \footnotesize $^a$ CFTC, University of Lisbon
    Av. Prof. Gama Pinto 2, 1649-003 Lisbon, Portugal}\\
  {\em \footnotesize $^b$ Ioffe Physico-Technical Institute, 194021
    St.Petersburg, Russia }\\  
  {\em \footnotesize $^c$ Dipartamento di Fisica and INFN, University of Rome III, Italy}\\
  {\em \footnotesize $^d$ St.Petersburg Institute of Nuclear Physics, 
    188300 Gatchina, Russia}
}  
\maketitle

\baselineskip=11.6pt

\begin{abstract}

The impact of the electroweak radiative corrections on the 
value of the forward-backward asymmetry in $e^+ e^-$ annihilation into a
quark-antiquark pair is considered in the double-logarithmic approximation
at energies much higher than the masses of the weak bosons.
\end{abstract}


\section{Introduction}

The forward-bacward asymmetry for the charged hadrons 
produced in $e^+e^-$ annihilation 
at high energies in vicinity of the $Z$ -boson mass has been the object 
of intensive theoretical and experimental investigation. It is  
interesting to make a theoretical analysis of this phenomenon for 
much higher enegies 
where the leading, double-logarithmic (DL) contributions come not only from 
integrating over 
virtual photon momenta but also from virtual $W$ and $Z$ bosons. 
Indeed, at 
the annihilation energies much higher than 100~Gev, when masses of  virtual 
electroweak bosons 
are small compared to their momenta, the initial $SU(2)$x$U(1)$ 
symmetry is restored in a certain sense and accounting for higher 
loop DL contributions involving the $W,Z$ bosons is nonless imporatnt than  
the standard accounting for the photon DL contributions. The value of the 
asymmetry at such energies would be expressed rather through 
 the Cazimir operators of the electroweak gauge group than through 
electric charges. In a sense, studing the forward-backward asymmetry at 
such enegries is one of the simplest ways to see the total effects of 
contributions of the electroweak radiative corrections of higher orders. 
Our theoretical study can be useful in future when  
Next linear colliders will explore $e^+e^-$ annihilation at very high
energies, probing further the Standard Model and eventually looking for
New Physics. 

In accordance with the present theoretical conceptions, we  divide 
investigation of the annihilation into two stages: first we calculate   
 the sub-process: 
$e^+e^-$ -annihilation into a quark-antiquark pair (which is 
 studied with perturbative methods) and then, numerically, account for 
hadronization effects which are quite different for converting the 
quarks into mesons and barions. We consider 
$e^+e^-$ annihilate into two hadronic jets in the kinematics when 
the leading particles 
of every jet goes in cmf close to the beam axis, so they are within the 
cones with opening angles $\theta \ll 1$ and the axes around the $e^-$ 
and  $e^+$ directions. We obtain that the forward-backward asymmetry 
manifests itself as follows: the number of the hadrons  
with the positive electric charges, $N_+$ 
in the cone around the  $e^+$ -direction 
exeeds the number of the negatively charged hadrons,  $N_-$  
(see Ref.~ \cite{egt}. It is depicted in Fig.~1. 
The opposite effect is true for the other cone, around the  $e^-$ direction. 
The space outside of the cones is neutral. The numerical 
evaluations of the asymmetry 
are plotted in Fig.~2 separately for the produced mesons and barions. 
In particular, Fig.2 shows that the value of the 
asymmetry is high enouph to be measured at energies $\approx 1$~TeV and 
grows steeply with energy. 

\begin{figure}[htbp]
\begin{center}
  \epsfig{file=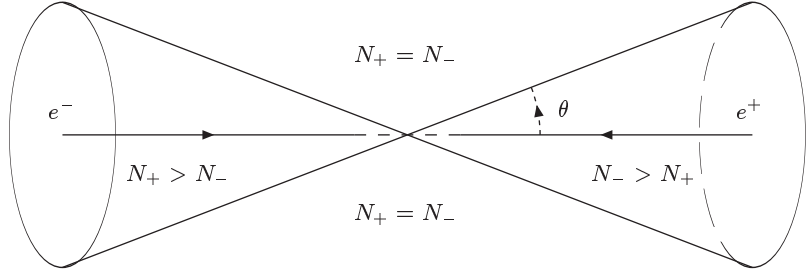}
  \caption{\it Relations between charged hadrons in diferent angular regions.
    \label{fig1}} 
\end{center}
\end{figure}

\begin{figure}[htbp]
\begin{center}
  \epsfig{file=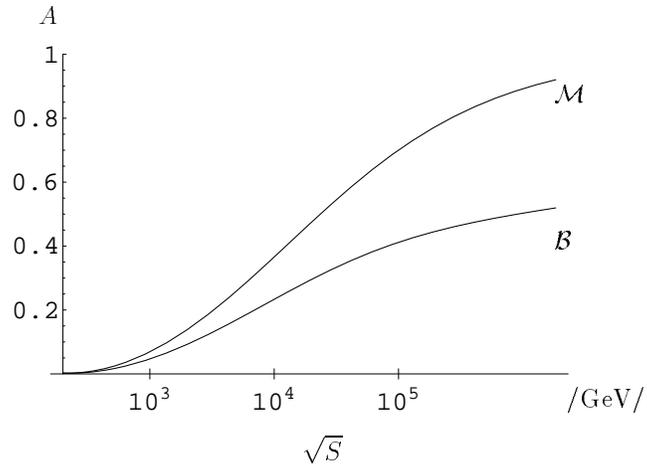}
  \caption{\it 
    Estimation of charge asymmetry $A$ of leading charged hadrons in
    $e^+e^-$ annihilation: The curve ${\cal M}$ is for the meson
    asymmetry and the curve ${\cal B}$ is for the asymmetry of
    barions.} 
  \label{fig2}
\end{center}
\end{figure}

In order to arrive at this result, let us consider first the 
basic sub-process of the annihilation: 
$e^+(p_2)e^-(p_1)$ annihilation into a quark $q(p_3)$ and its antiparticle 
$\bar{q}$. All these particles can belong either to the $SU(2)$ doublets or to 
the singlets. Our first goal is to calculate the scattering amplitudes of 
the annihilation in the $t$- kinematics where 
\begin{equation}
\label{tkin}
s = (p_1 + p_2)^2 \gg t =(p_3 - p_1)^2  
\end{equation}   
and in the $u$- kinematics: 
\begin{equation}
\label{ukin}
s = (p_1 + p_2)^2 \gg u =(p_4 - p_1)^2 ~.  
\end{equation}   

In order to account for DL contributions to all orders in the electroweak 
couplings, we use the evolution equations with respect to the infrared 
cut-off. This cut-off $M$ is chosen in the transverse momentum space so 
that momenta of all virtual particles obey 
\begin{equation}
\label{m}
k_{i \perp} > M ~.
\end{equation}

Although only the Feynman graphs vith virtual photons can have the 
infrared divergencies, it is convenient to keep the 
restriction~(\ref{m}) for momenta of all virtual particles, assuming that 
\begin{equation}
\label{mwz}
M \geq M_Z \approx M_W ~.
\end{equation}       

With assumptions of Eqs.~(\ref{m}\ref{mwz}), one can neglect all masses and 
be safe of the infrared singularities at the same moment. On the other hand, 
the scattering amplitudes now depend on $M$. It makes possible to evolute 
them in $M$ and to put $M = M_Z \approx M_W$ in the final expressions. 
As DL contributions appear in the regions where $k_{i \perp}$ obey the 
strong inequalities of the kind  $k_{i \perp} \gg k_{j \perp}$, it is 
always possible to find the virtual particle with minimal  
$(\equiv k_{ \perp})$ transverse momentum in 
every such a region. Obviously, only integration over $k_{ \perp}$ 
involves $M$ as the lowest limit. Integrations over other transverse 
momenta are $M$-independent. 
DL contributions of the softest particles can be factorized. It allows to 
compose infrared evolution equations (IREE) for the scattering amplitudes. 
The most difficult is the case when both the initial electron and the final 
quark belong to the $SU(2)$ doublets. In order to simplify the IREE, 
one can use the $SU(2)$ simmetry restored at such high energies and consider 
annihilation of lepton-antilepton pair into a quark-antiquark pair. 
After that, it is convenient to exapand the scattering amplitude into 
the sum of the irreducible $SU(2)$ representations, using the standard 
projection operators multiplied by the invariant amplitudes 
$A_j, (j= 1,2,3,4)$.    
At last, in order to calculate the invariant amplitudes in kinematics 
(\ref{tkin}, \ref{ukin}), it is convenient 
to use the Mellin transform     
\begin{equation}
\label{mellin}
A_j =
\int_{-\imath \infty}^{\imath \infty}
\frac{d\omega}{2\pi\imath} \left(\frac{s}{\kappa}\right)^{\omega}\,
F_j(\omega)
\end{equation}
where $\kappa = t, j= 1,2$ for  $A_j$ in kinematics (\ref{tkin}) and 
 $\kappa = u, j= 3,4$ when the kinematics is  the $u$ -inematics of 
Eq.~(\ref{tkin}).  In the case of the collinear kinematics where 
$\kappa = M^2$, amplitudes $F_j$ obey (we consider 
amplitudes with the positive signatures only):  
\begin{equation}
\label{eqfj}
\omega F_j(\omega) = a_j +
\frac{b_j}{8\pi^2} \frac{dF^{(+)}_j(\omega)}{d \omega} +
\frac{c_j}{8\pi^2} \left[F^{(+)}_j(\omega)\right]^2 ~,
\end{equation}
with $a_j, b_j, c_j$ being numerical factors: 
\begin{eqnarray}
\label{aj}
a_1 = \frac{3g^2+{g'}^2 Y_1Y_2}{4}, 
~~a_2 = \frac{-g^2+{g'}^2 Y_1Y_2}{4},           \\ \nonumber
~~a_3=\frac{- 3g^2 + {g'}^2 Y_1Y_2}{4},
~~a_4 = \frac{g^2 + {g'}^2 Y_1Y_2}{4}~~,
\end{eqnarray} 

\begin{eqnarray}
\label{bj}
b_1 = \frac{{g'}^2 (Y_1 - Y_2)^2}{4}, 
~~b_2 = \frac{8g^2+{g'}^2 (Y_1 - Y_2)^2}{4},           \\ \nonumber
~~b_3=\frac{{g'}^2 (Y_1 - Y_2)^2}{4},
~~b_4 = \frac{8g^2 + {g'}^2 (Y_1 + Y_2)^2}{4}~~,
\end{eqnarray}

\begin{equation}
\label{cj}
c_1 = c_2 = 1, ~~~c_3 = c_4 = -1~.
\end{equation}

Solutions to Eq.~(\ref{eqfj}) can be expressed
in terms of the Parabolic cylinder functions $D_p$:

\begin{equation}
\label{fj}
F_j(\omega) = \frac{a_j}{\lambda_j}
\frac{D_{p_j - 1} (\omega/\lambda_j)}{D_{p_j} (\omega/\lambda_j)}~
\end{equation}
where $p_j = a_jc_j/b_j$ and $\lambda_j = \sqrt{b_j/(8\pi^2)}$. 

When the scattering angles are becoming larger so that 

\begin{equation}
\label{bigkappa}
s \gg -\kappa \gg \mu^2 ~,
\end{equation}
the invariant amplitudes $A_j$ are expressed in terms of $F_j$ in the 
following way: 

\begin{equation}
\label{mpostu}
A_j(\rho,\kappa) = a_j\, S_j\,  
\int_{-\imath \infty}^{\imath \infty} \frac{dl}{2\pi\imath}\,
e^{\lambda_j l (\rho-\eta)}
\frac{D_{p_j-1}(l+\lambda_j\eta)}
{D_{p_j}(l+\lambda_j\eta)}
\end{equation}
where $\eta=\ln(\kappa/\mu^2)$ and factors $h_j$ can be taken from  
Ref.~\cite{egt}:

\begin{eqnarray}
\label{hj}
h_1 = \frac{3 g^2 + {g'}^2 Y_l Y_q}{2} ,\qquad
h_2 = \frac{-g^2 + {g'}^2 Y_l Y_q}{2} ,  \\  \nonumber
h_3 = \frac{3 g^2 - {g'}^2 Y_l Y_q}{2} ,\qquad
h_4 = \frac{-g^2 - {g'}^2 Y_l Y_q}{2} .
\end{eqnarray}

The factor $S_j$ in Eq.~(\ref{mpostu}) is
the Sudakov form factor. Actually it is a product of the Sudakov form
factors of the leptons and of the 
quarks.  As follows from Eqs.~(\ref{bj},\ref{hj}), 
Due to the gauge invariance, it does not depend on
$j$, i.e.  is same for all invariant amplitudes. It can be seen also 
explicitly from Eqs.~(\ref{bj},\ref{hj}), 
 
\begin{equation}
\label{sudakew}
S= \exp\left[-\,\frac{1}{8\pi^2} \,
\left(\frac32\, g^2 + \frac{Y_l^2+Y_q^2}{4}\, {g'}^2 \right)
\, \frac{{\eta'}^2}{2}\right] ~.
\end{equation}

This form factor accumulates DL contributions of soft virtual EW bosons and
vanishes in the final expressions for the cross sections when
bremsstrahlung of
soft EW bosons with the cmf energies $\epsilon_k$,  
$\mu < \epsilon_k < \sqrt{\kappa}$, is  taken into account. 

According to the results of Ref.~\cite{egt},  
 the amplitude for
the forward $e^+e^- \to u\bar{u}$ -annihilation, $M^F_u$ is expressed in
terms of amplitudes $A_3, A_4$ of Eq.~(\ref{mpostu}):

\begin{equation}
\label{mfu}
M_u^F =   \frac{A_3 + A_4}{2} ~,
\end{equation}
whereas the backward amplitude $M_u^B$
for the same quarks is equal to the amplitude $A_4$ of
Eq.~(\ref{mpostu}):

\begin{equation}
\label{mbu}
M_u^B = A_4  ~.
\end{equation}

Similarly, the forward amplitude $M^F_d$ for
$e^+e^- \to d\bar{d}$ is

\begin{equation}
\label{mfd}
M_d^F =  \frac{A_1 + A_2}{2} ~,
\end{equation}
while  the backward amplitude for this process is

\begin{equation}
\label{mbd}
M_d^B = A_2 ~.
\end{equation}

We use the folloing terminology: 
by the forward kinematics for $e^+e^-\to q\bar{q}$ -annihilation
we mean that quarks with positive electric
charges, $u$ and $\bar{d}$, are produced around the initial $e^+$ - beam,
in the cmf,
within a cone with a small opening angle $\theta$,

\begin{equation}
\label{thetatu}
1\gg\theta\ge \theta_0 = \frac{2M}{\sqrt{s}}~
\end{equation}

By backward kinematics we means just the opposite -- the electric charge
scatters backwards in a cone with the same opening angles.

The differential cross section $d\sigma_F$ for the forward annihilation is

\begin{equation}
\label{sigmaf}
d\sigma_F = d\sigma^{(0)}\,\left[|M_u^F|^2 + |M_d^F|^2\right] \equiv
d\sigma^{(0)}\,\left[F_u + F_d\right]  ~,
\end{equation}
and similarly, the differential cross section $d\sigma_B$ for the
backward annihilation is

\begin{equation}
\label{sigmab}
d\sigma_B = d\sigma^{(0)}\, \left[|M_u^B|^2 + |M_d^B|^2\right] \equiv
d\sigma^{(0)}\, \left[B_u + B_d\right],
\end{equation}
where $d\sigma^{(0)}$ stands
for the Born cross section, though without couplings.We define the
forward-backward asymmetry as:

\begin{equation}
\label{defasym}
A \equiv \frac{d\sigma_F - d\sigma_B }{d\sigma_F + d\sigma_B}
= \frac{F - B}{F + B} ~
\end{equation}
where

\begin{equation}
\label{FB}
F = F_u + F_d, \qquad  B = B_u + B_d ~.
\end{equation}


Contributions to the asymmetry from right leptons and quarks can be 
ealily obtained in a similar way. Then, using the standard programmes for 
hadronisation\cite{s} and doing numerical calculations, we obtain the 
forward-backward asymmetry for $e^+e^-$ annihilation into charged mesons 
and barions. The results are plotted in Fig.~2.  

Finally, we would like to note that when  the annihilation 
energy is high enouph to produce, in addition to 
the two quark jets, $W$ or $Z$ bosons with energies $\gg$ 100~GeV, it 
turnes out that  the cross section  
$\sigma^{(n \gamma)}$ of $n$ hard photon production, the cross section 
 $\sigma^{(n Z)}$ of $n Z$ boson production and the cross section 
 $\sigma^{(n W)}$ of $n W$ boson production    
obey very simple asymptotical relations (see Ref.~ \cite{eot}:       

\begin{equation}
\label{gammazn}
\frac{\sigma^{(nZ)}}{\sigma^{(n\gamma)}} \approx \tan^{2n} \theta_W ~,
\end{equation}
\begin{equation}
\label{gammawn}
\frac{\sigma^{(n \gamma)}}{\sigma^{(n W)}} \sim s^{-0.36}.
\end{equation}

The ratios of these cross sections as functions of the annihilation 
energy are plotted in Figs.~3,4.

\begin{figure}[htbp]
  \begin{center}
  \epsfig{file=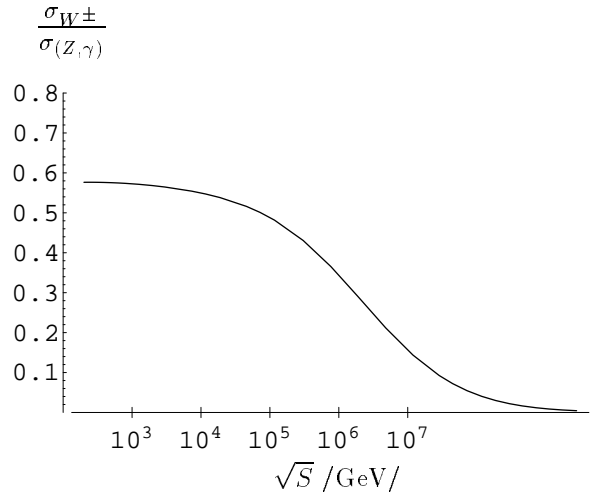}
  \caption{\it
    Total energy dependence of $W^{\pm}$ to
    $(Z,\gamma)$ rate in $e^+e^-$ annihilation.
    \label{exfig3} }
\end{center}
\end{figure}
\begin{figure}[htbp]
  \begin{center}
  \epsfig{file=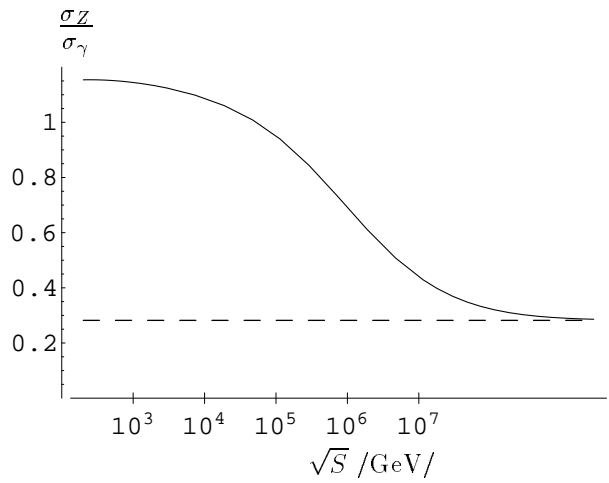}
  \caption{\it
    Total energy dependence of $Z$ to
    $\gamma$ rate in $e^+e^-$ annihilation. The dashed line shows the
    asymptotical value of the ratio: $\tan^2\theta_W\approx 0.28$~. 
    \label{exfig4} } 
\end{center}
\end{figure}

\section{Acknowledgement}
The work is supported by grants POCTI/FNU/49523/2002, SFRH/BD/6455/2001   
and RSGSS-1124.2003.2 .


\begin{thebibliography}{99}

\bibitem{egt} B.I.~Ermolaev {\it et al}, 
Phys.Rev. {\bf D67}, 014017, (2003).

\bibitem{eot} B.I.~Ermolaev {\it et al}, 
Phys.Rev. {\bf D66}, 114018, (2002).

\bibitem{s}T.Sj\"ostrand. Computer Physics Commun. {\bf 82}, 74, (1994).  

\end{thebibliography}
\end{document}